\documentclass[9pt,twocolumn,twoside]{opticajnl}
\journal{opticajournal} 

\setboolean{shortarticle}{false}

\usepackage{lineno}
\usepackage{fancyhdr}

\usepackage[absolute,overlay]{textpos}

\newcommand{\neff}{n_\text{eff}}

\newcommand{\pd}{\partial}

\newcommand{\mum}{~\mu\text{m}}

\newcommand{\nappr}{\neff^\text{(appr.)}}

\title{Analytical Expressions for Effective Indices of Modes of Optical Fibers Near and Beyond Cutoff}

\author[1,*]{Aku Antikainen}
\author[1,2]{Robert W. Boyd}

\affil[1]{Institute of Optics, University of Rochester, Rochester, NY 14627, USA}
\affil[2]{Department of Physics, University of Ottawa, Ottawa, Ontario K1N 6N5, Canada}

\affil[*]{aku.antikainen@rochester.edu}

\begin{abstract}
We derive an analytical expression for the effective indices of modes of circular step-index fibers valid near their cutoff wavelengths. The approximation, being a first-order Taylor series of a smooth function, is also valid for the real part of the effective index beyond cutoff where the modes become lossy. The approximation is used to derive certain previously unknown mode properties. For example, it is shown that for non-dispersive materials the EH-mode group index at cutoff, surprisingly, does not depend on wavelength, core radius, or even radial mode order.
\end{abstract}

\setboolean{displaycopyright}{false} 

\begin{document}

\begin{textblock*}{167.6mm}(25.4mm, 16mm) 
{\color{red}
   \noindent \textbf{Accepted for publication in Journal of Lightwave Technology (https://doi.org/10.1109/jlt.2024.3513437)}
	\newline
	\newline
   \noindent \textbf{0733-8724 \copyright  2024 IEEE. All rights reserved, including rights for text and data mining, and training of artificial intelligence and similar technologies. Personal use is permitted, but republication/redistribution requires IEEE permission. See https://www.ieee.org/publications/rights/index.html for more information. }}
\end{textblock*}

\maketitle

\section{Introduction}

Circular step-index fibers are the simplest type of optical fiber, and they consist of a circular core of high-index medium surrounded by a cladding of lower refractive index. Whereas graded-index fibers have dominated the multimode fiber scene in recent decades, step-index fibers have been shown to support optical phenomena not possible in graded-index fibers. A notable example of such a phenomenon is soliton self-mode conversion, in which a soliton spontaneously changes color and switches from one spatial mode to another \cite{Antikainen2019, Rishoej2019, Kabagoez2021a}, making step-index multimode fibers useful for the generation of energetic dual- or multicolor \cite{Kabagoez2021} ultrashort optical pulses at unconventional wavelengths, for example.

Linear light propagation in step-index fibers is determined by its color and its spatial mode content, with different spatial modes having different effective refractive indices. The study and modelling of light in multi-mode fibers thus requires determination of the effective indices of the modes. The determination of modes of circular step-index fibers can be done semi-analytically by exploiting the fact that the radial dependence of the electric and magnetic fields making up the modes are described by Bessel functions. Even still, the determination of modal parameters becomes computationally intensive and oftentimes also numerically inaccurate for heavily multimode fibers. Furthermore, modelling pulse propagation in fibers requires knowledge on the effective index for a range of wavelengths, thus increasing the number of times the effective index has to be computed. Therefore, tools to speed up the calculations can be invaluable, especially when looking for fiber designs with specific properties.

Any mode except for the fundamental fiber mode has a cutoff wavelength, and the mode becomes lossy for wavelengths longer than the cutoff wavelength. Near cutoff, the dispersion properties of the modes can change drastically with wavelength, and higher-order modes can be used for dispersion compensation in this wavelength regime \cite{Poole1994}. It has recently been shown \cite{Ma2023}, that for modes with high azimuthal order (orbital angular momentum, OAM) and low radial order, the losses beyond cutoff can be small enough to allow these leaky modes to be useful for pulse transmission. Furthermore, these leaky modes are much more stable against fiber imperfections, as mode mixing is mitigated due to the fact that high radial orders experience larger losses beyond cutoff, making random coupling from low radial orders to higher orders negligible. 

The behavior of modes near the cutoff wavelength is therefore of interest not just because of the exotic dispersive properties but also due to the fact that the properties of high-OAM modes are more desirable beyond cutoff than those of conventional modes below the cutoff wavelength. Here, we derive a linear approximation to the effective indices of modes near their cutoff wavelengths, simply expressing the effective index as a function of wavelength and fiber parameters. The approximation can be used to speed up the determination of effective indices, remains valid beyond cutoff, and can be used to determine mode properties at cutoff analytically. As an example, we derive certain asymptotic formulae for mode behavior when their radial and/or azimuthal order is increased, as well as a curious result showing that the group index of EH-modes at cutoff is independent of waveguide dispersion, i.e., in the absence of material dispersion, it does not depend on fiber core size, the cutoff wavelength, or even radial mode order.

\section{First-Order Approximation to the Modal Determinant Equation}

Let the core radius be $a$, core index $n_1$, and cladding index $n_2$. We focus on the effects of waveguide dispersion, and material dispersion is neglected by assuming that the indices $n_1$ and $n_2$ are constants. The cladding is assumed to be infinite, which is a valid approximation for a plethora of cases of interest. The validity of the approximation essentially requires that the cladding is thick enough such that the mode fields have decayed to negligible levels by the edge of the cladding. Obviously, the thicker the cladding, the better the approximation, but what constitutes as ``thick enough'' also depends on the mode orders: modes of high radial order and/or low azimuthal order decay slower away from the core and therefore require physically thicker claddings. Such modes are thus more prone to become cladding-guided beyond cutoff, in which case the whole fiber essentially acts as the ``core'' in the mathematical sense, and the medium surrounding the cladding (generally a coating and/or air) as the ``cladding''. All modes will eventually become cladding-guided for long enough wavelengths, but the infinite cladding remains a standard textbook approximation and works especially well for modes of high azimuthal order due to the rapid field decay in the radial direction outside of the core, even beyond cutoff \cite{Ma2023}.

Let us define the usual normalized frequency as $V = a k_0 \sqrt{n_1^2 - n_2^2}$, where $k_0 = 2 \pi / \lambda_0$ is the vacuum wavenumber. In cylindrical coordinates $r, \phi, z$, the $z$-components of the electric and magnetic fields are of the functional form
\begin{align}
    E_z(r, \phi, z,t) &= A J_m(pr) e^{i m \phi} e^{i(\beta z - \omega t)}, ~~ r \leq a \\
    H_z(r, \phi, z,t) &= B J_m(pr) e^{i m \phi} e^{i(\beta z - \omega t)}, ~~ r \leq a \\
    E_z(r, \phi, z,t) &= C K_m(qr) e^{i m \phi} e^{i(\beta z - \omega t)}, ~~ r \geq a \\
    H_z(r, \phi, z,t) &= D K_m(qr) e^{i m \phi} e^{i(\beta z - \omega t)}, ~~ r \geq a 
\end{align}
where $A, B, C, D, m, \beta, \omega$ are constants, $m$ being the azimuthal order, $\beta = \neff k_0 = 2 \pi \neff / \lambda_0$ the propagation constant, and $\omega$ the angular frequency. $J_m$ is the Bessel function of the first kind, and $K_m$ is the modified Bessel function of the second kind. The functional form is dictated by Maxwell's equations and the cylindrical symmetry. The other field components, $E_\phi, E_r, H_\phi, H_r$, can be obtained from $E_z$ and $H_z$. The azimuthal components $E_\phi, E_z, H_\phi, H_z$ need to be continuous across the core-cladding interface, which then ties the constants  $A, B, C,$ and $D$ to one another. In matrix form, these continuity conditions can be written as
\begin{equation}
    M \begin{bmatrix}
        A \\ B \\ C \\ D
    \end{bmatrix}
    =
        \begin{bmatrix}
        0 \\ 0 \\ 0 \\ 0 
    \end{bmatrix} , \label{eq:mateq}
\end{equation}
where 
\begin{equation}
    M 
    =
    \begin{bmatrix}
    J_m & 0 & -K_m & 0 \\
    0 & J_m & 0 & -K_m  \\
 i \frac{m \beta}{a p^2} J_m  
 & 		
 - \frac{\omega \mu_1}{p} J_m'  
 & 
i \frac{m \beta}{a q^2} K_m
 & 
- \frac{\omega \mu_2}{q} K_m' 
\\
\frac{\omega \varepsilon_1}{p} J_m' 
&
i \frac{m \beta}{a p^2} J_m
& 
\frac{\omega \varepsilon_2}{q} K_m'
&
i \frac{m \beta}{a q^2} K_m
    \end{bmatrix},
\end{equation}
where
\begin{align}
    p &= k_0 \sqrt{n_1^2 - \neff^2} \\
    q &= k_0 \sqrt{\neff^2 - n_2^2} ,
\end{align}
and $J_m = J_m(a p)$, $J_m' = J_m'(a p)$, $K_m = K_m(a q)$, and $K_m' = K_m'(a q)$.
Equation~(\ref{eq:mateq}) has a non-trivial solution if and only if $\det(M) = 0$. The propagation constants $\beta$ can then be determined from this condition numerically.

All modes except for the fundamental mode have a cutoff wavelength, at which the effective index $\neff$ becomes equal to the cladding index and beyond which the mode becomes lossy. The effective index is a smooth function of the wavelength, and hence near the cutoff wavelength the effective index can be written as $\neff = n_2 + d$, where $d$ is small. This can then be plugged into the equation $\det(M) = 0$. Since $d$ is small, the determinant is well approximated by a series expansion in $d$ centered around $d = 0$. The Bessel functions $K_m$ and $K_m'$ and the factor $1/q$ diverge at $d = 0$, so the series expansion will not be a Taylor series, but a Puiseux series. The Bessel functions $J_m$ and $J_m'$ have simple Taylor expansions at $d = 0$, but for $K_m$ the following series needs to be used:
\begin{align}
    K_m(qa) = \frac{1}{(a q)^m} \left[ D_m + F_m (a q)^2 + G_m (a q)^4 + \mathcal{O}(d^{3 - \delta}) \right] , \label{eq:Kappr}
\end{align}
where
\begin{align}
    D_m &=   2^{m - 1} (m - 1)!  \label{eq:Dm}  \\ 
    F_m &= - 2^{m - 3} (m - 2)!  \label{eq:Fm}  \\ 
    G_m &=   2^{m - 6} (m - 3)!  \label{eq:Gm}  
\end{align}
and $\delta$ is an arbitrarily small positive real number. The expansion is valid for $m \geq 3$. For $m \geq 4$ we can set $\delta = 0$, but for $m = 3$ it is needed to account for an error term proportional to $d^3 \ln d$. A similar expansion is valid for $m \leq 2$, but the factors $D_m, F_m,$ and $G_m$ will depend on $\ln(aq)$. The same principles can be applied to modes with $m \leq 2$, but the logarithm terms complicate the mathematics, leading to having to express $d$ through the Lambert $W$-function in the end. In what follows, it will therefore be assumed $m \geq 3$. We can now plug in Eq.~(\ref{eq:Kappr}) into $\det(M) = 0$ and expand all the other factors in the determinant as a series in $d$ as well. When keeping only the two dominant order terms in $d$, the modal equation can be written in the form
\begin{equation}
    0 = (a q)^N \det(M) \approx a_{M,m} + b_{M,m} d ,
\end{equation}
where $N$ is an integer, $a_{M,m}$ and $b_{M,m}$ are constants that depend on the wavelength and fiber parameters. The equation is trivially solved for $d$, yielding an approximation for $\neff$ valid near the cutoff frequency. The full derivation can be found in the supplement, but the final result is
\begin{align}
    \nappr = n_2 - \frac{m J_m}{n_2} \frac{  V n_2^2 J_m  - (m - 1) (n_1^2 + n_2^2) J_{m-1}}
    { S_1 J_{m+1}^2 +  S_2 V J_{m+1} J_m +  S_3 J_{m}^2 } V^3 \label{eq:neffappr} ,
\end{align}
where
\begin{align}
    S_1 &= (m + 2)(m-1)  V^4  + 2 f (m^2 - 1) V^2 \\
    S_2 &= - (m - 1) V^4 - 2 (2 m^3 - m^2 - m - f) V^2 \nonumber \\ 
		    &- 8 f m^2 (m - 1)  \\
    S_3 &=  \left[ m^2 - \frac{m - 1}{m - 2} f \right] V^4 + 4 m^2 (m - 1)^2 V^2 \nonumber \\
		    &+ 8 f m^2 (m - 1)^2 \\
    f &= a^2 k_0^2 n_2^2 ,
\end{align}
and the Bessel functions $J_m$ and $J_{m-1}$ are evaluated at $V$. Note that Eq.~(\ref{eq:neffappr}) contains the cutoff conditions for HE- and EH-modes: $\nappr = n_2$ if and only if 
\begin{equation}
    J_m(V) = 0 \label{eq:EHcutoff} ,
\end{equation}
which is the cutoff condition for EH-modes, or if
\begin{equation}
    V n_2^2 J_m  - (m - 1) (n_1^2 + n_2^2) J_{m-1} = 0 \label{eq:HEcutoff} ,
\end{equation}
which is the cutoff condition for HE-modes. Let us denote the solutions of Eqs.~(\ref{eq:EHcutoff}) and (\ref{eq:HEcutoff}) in ascending order by $j_{mn}$ and $s_{mn}$, respectively, so that $j_{m1}$ is the smallest positive zero of $J_m$, for example.

\section{Behavior of Modes Near Cutoff}

The expression for the approximate effective index in Eq.~(\ref{eq:neffappr}) is evidently very nonlinear in the vacuum wavelength $\lambda_0$, but, being a first-order expansion in $d$, gives an approximation that is locally co-linear to the actual effective index with respect to wavelength at cutoff. Equation.~(\ref{eq:neffappr}) can therefore be utilized to further derive approximations of the form
\begin{equation}
    \neff \approx n_2 + \kappa (\lambda_0 - \lambda_c) ,
\end{equation}
where $\kappa$ is a constant and $\lambda_c$ is a cutoff wavelength. The cutoff wavelengths are related to $j_{mn}$ and $s_{mn}$ as 
\begin{equation}
    \lambda_c = \frac{2 \pi a}{x_{mn}} \sqrt{n_1^2 - n_2^2} ,
\end{equation}
where $x_{mn} = j_{mn}$ for EH-modes and $x_{mn} = s_{mn}$ for HE-modes. Again, the derivation of the following results can be found in the supplement, but the approximate effective indices for HE-modes can be written as 
\begin{align}
    \neff^\text{HE} - n_2 \approx 
    &\left( 1 - \frac{\lambda_0}{\lambda_c}  \right) \frac{m (n_1^4 - n_2^4)}{n_2}  \\
    &\times \frac{(m-1)^2 (n_1^4 - n_2^4) + n_2^4 s_{mn}^2}{ (n_1^2 - n_2^2) P_m + n_2^2 Q_m s_{mn}^2 }  \label{eq:HEneff} ,
\end{align}
where
\begin{align}
    P_m &= m (m-1) (m-2) n_1^2 + m^2 (m-1) n_2^2 \\
    Q_m &= \frac{m-1}{m-2} (n_1^2 + n_2^2)^2  + (m-2) n_2^2 (n_1^2 + n_2^2) + 2 n_2^4 .
\end{align}
The expression for EH-modes is simpler:
\begin{align}
    \neff^\text{EH} - n_2
    &\approx  \left( 1 - \frac{\lambda_0}{\lambda_c}  \right)
              \frac{m (n_1^4 - n_2^4) }{n_2 [  (m + 2) n_1^2 + m n_2^2 ]} \label{eq:EHneff} .
\end{align}
Equations~(\ref{eq:HEneff}) and (\ref{eq:EHneff}) are linear approximations to the effective index at the cutoff wavelength. They can therefore be used to analytically determine the \emph{exact} group index $n_g$ at any cutoff wavelength through
\begin{equation}
    n_g(\lambda_c) = n_2 - \lambda_c \left[ \frac{d n(\lambda_0)}{d \lambda_0} \right]_{\lambda_0 = \lambda_c} .
\end{equation}
This yields
\begin{equation}
    n_g^\text{HE} = n_2 + \frac{m (n_1^4 - n_2^4)}{n_2}
\frac{(m-1)^2 (n_1^4 - n_2^4) + n_2^4 s_{mn}^2}
    {       
    (n_1^2 - n_2^2) P_m + n_2^2 Q_m s_{mn}^2 .
    }  , \label{eq:HEng} 
\end{equation}
and
\begin{equation}
    n_g^\text{EH} = n_2 + \frac{m (n_1^4 - n_2^4)}{n_2 [  (m + 2) n_1^2 + m n_2^2 ]} .
    \label{eq:EHng}
\end{equation}
A surprising result can be seen from Eq.~(\ref{eq:EHng}): \emph{The group velocity of an EH mode at the cutoff wavelength does not depend on the cutoff wavelength, core radius, or radial mode order}. As a reminder, material dispersion has been neglected in the derivation, and the result of Eq.~(\ref{eq:EHng}) essentially means that the EH mode group velocity at cutoff is unaffected by waveguide dispersion. Since Eqs.~(\ref{eq:HEneff}) and (\ref{eq:EHneff}) are first-order approximations to the effective index, the inclusion of material dispersion would simply add terms proportional to the wavelength slopes $d n_1 / d \lambda$ and $d n_2 / d \lambda$ to them.

Equations~(\ref{eq:HEneff}) and (\ref{eq:EHneff}) can also be used to determine mode behavior near cutoff when the azimuthal and/or radial mode order is increased. The azimuthal mode order is arguable the more interesting one, since it has been demonstrated that modes of high azimuthal order can propagate with small loss beyond cutoff. Equations~(\ref{eq:HEneff}) and (\ref{eq:EHneff}) work beyond the cutoff wavelength as well, since they are simply linear approximations to a smooth function, and they give the approximate real part of the effective index on both sides of the cutoff wavelength. In the mathematical sense, modes exist beyond cutoff, but they would require infinite energy, which is why they become lossy in real life.

Consider, first, the simpler EH-modes for large azimuthal order $m$ and of radial order one. The cutoff wavelengths of these modes are related to $J_{m1}$, and $j_{m1}$ obey inequalities given by \cite{Giordano1983}
\begin{equation}
    m +  k m^{1/3} + \frac{0.5}{m^{1/3}} \geq j_{m1} \geq m + k m^{1/3} + \frac{1.357}{m^{1/3}}
\end{equation}
for $m \geq 1$, where $k = 1.855757\ldots$. Plugging these into Eq.~(\ref{eq:EHneff}) then gives upper and lower bounds for the effective indices near the cutoff. The upper bound is particularly useful when determining the effective indices numerically, as below the cutoff wavelength the effective index is naturally bound by $n_2$ from below. The inequalities for $j_{mn}$ also show that
\begin{equation}
    \neff^\text{EH} 
    \rightarrow
    \frac{n_1^2}{n_2} - \lambda_0 \frac{m +  1.855757 m^{1/3}}{2 \pi a n_2} \sqrt{n_1^2 - n_2^2}
\end{equation}
as $m \rightarrow \infty$. This gives the asymptotic behavior of EH-modes near cutoff for large orbital angular momenta.

As for the HE modes, first note that the cutoff condition can be written as 
\begin{equation}
     \frac{n_1^2 + n_2^2}{n_1^2 - n_2^2} J_{m-2}(V_c) = J_m(V_c) \label{eq:HEcutoff2} .
\end{equation}
The zeros of $J_{m-2}$ and $J_m$ coalesce (i.e. $j_{(m-2)n} - j_{m(n-1)} \rightarrow 0$) as $n \rightarrow \infty$, which means $s_{mn} \rightarrow j_{(m-2)n}$ as $n \rightarrow \infty$. Furthermore, for most fibers, the factor in front of $J_{m-2}(V_c)$ is very large, which means that the solutions $s_{mn}$ are close to $j_{(m-2)n}$ even for small radial orders $n$. We can therefore make the approximation
\begin{equation}
    s_{m1} \approx m - 2 + 1.855757 (m - 2)^{1/3} .
\end{equation}
Plugging this approximation into Eq.~(\ref{eq:HEneff}) gives
\begin{align}
    \neff^\text{HE} - n_2 \approx 
    &\left( 1 - \frac{m - 2 + 1.855757 (m - 2)^{1/3}}{2 \pi a \sqrt{n_1^2 - n_2^2}} \lambda_0  \right) \\
    &\times \frac{n_1^4}{n_2} \frac{n_1^4 - n_2^4}{n_1^4 + n_1^2 n_2^4 + n_2^6 - n_2^4}  \label{eq:HEneff} , \nonumber
\end{align}
as $m \rightarrow \infty$. The weak guidance approximation ($n_1 \approx n_2$) could be dropped with a better approximation for $s_{mn}$ in terms of $m$, $n_1$, and $n_2$.

Zeros of $J_m$ far on the positive real axis also have an asymptotic approximation \cite{McMahon1894}:
\begin{equation}
    j_{mn} = \left( n + \frac{m}{2} - \frac{1}{4} \right) \pi - \frac{4 m^2 - 1}{\left( 8 n + 4 m - 2 \right) \pi} + \mathcal{O}(n^{-3}) \label{eq:Jmzeroappr}
\end{equation}
that can be plugged into Eqs.~(\ref{eq:HEneff}) and (\ref{eq:EHneff}) to see how the behavior of effective index near cutoff changes with radial order. This approximation is also useful in determining the cutoff wavelengths themselves, as it provides a good starting guess for any numerical root-finding algorithm. Equation~(\ref{eq:Jmzeroappr}) shows that for large radial orders, the wavelength-slope of the effective index increases slightly faster than but approximately linearly with radial mode order.

\section{Discussion}

\begin{figure}[ht]
\centering
\includegraphics[width=\linewidth]{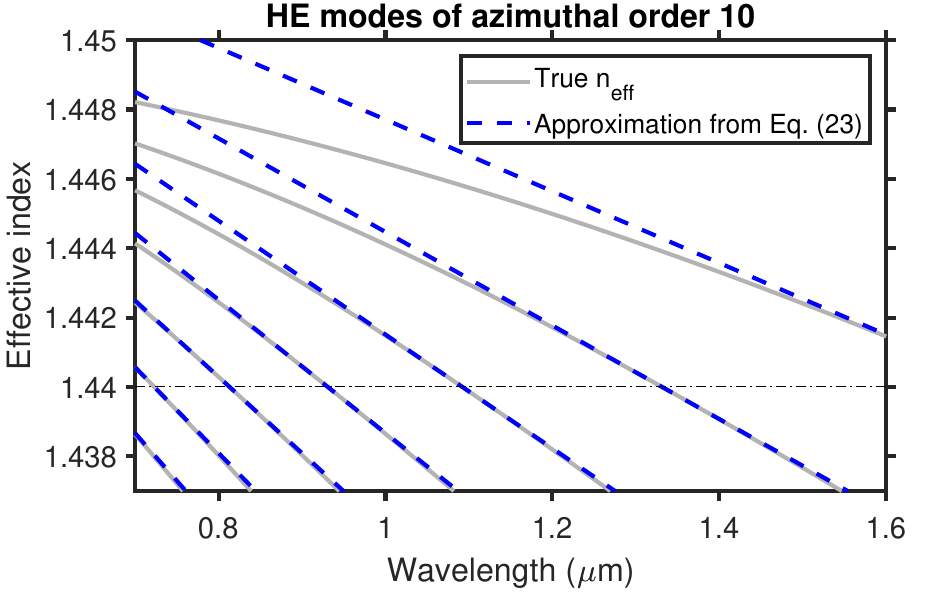}
\caption{The effective index of HE-modes of azimuthal order $m = 10$ in the example fiber. The radial mode order starts at one at the top (highest effective index) and increases from top to bottom. The solid gray lines show the actual effective indices, and the dashed blue lines are the linear approximations from Eq.~(\ref{eq:HEneff}). The horizontal dot-dash line indicates the cladding index, below which the modes are lossy.}
\label{fig1}
\end{figure}

As an example, consider a fiber with core radius $a = 20 \mum$, core index $n_1 = 1.45$ and cladding index $1.44$. Figure~\ref{fig1} shows the effective indices of the supported HE modes of azimuthal order $m = 10$. The lowest radial order is $1$ and the corresponding mode has the highest effective index.

\begin{figure}[ht]
\centering
\includegraphics[width=\linewidth]{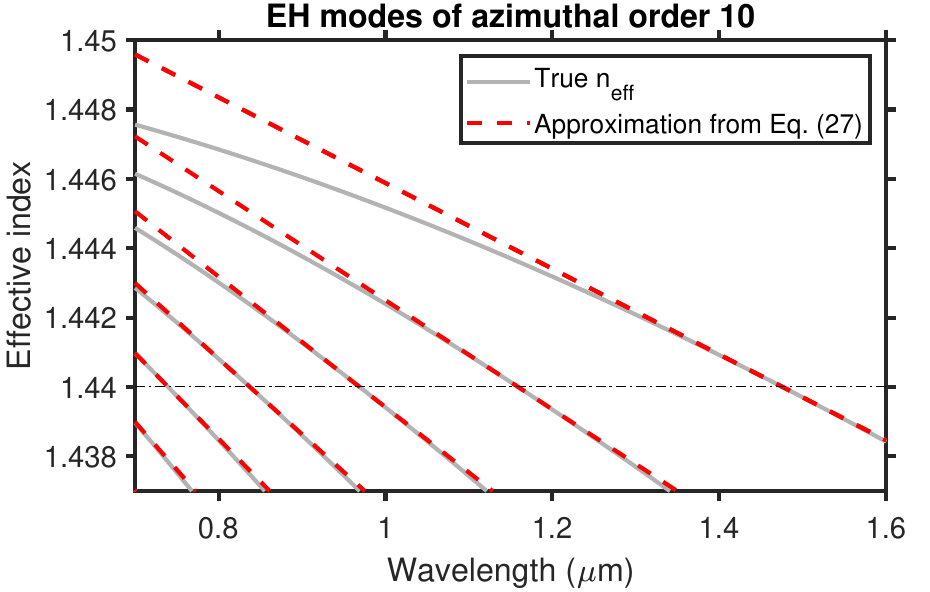}
\caption{The effective index of EH-modes with $m = 10$ in the example fiber. The solid gray lines show the true effective indices. The dashed red lines show the approximate effective index using  Eq.~(\ref{eq:HEneff}).}
\label{fig2}
\end{figure}

The linear approximation works very well near the cutoff wavelengths, as can be seen in the figure. The same is true for EH-modes utilising Eq.~(\ref{eq:EHneff}). This case is shown in Fig.~(\ref{fig2}).

\begin{figure}[ht]
\centering
\includegraphics[width=\linewidth]{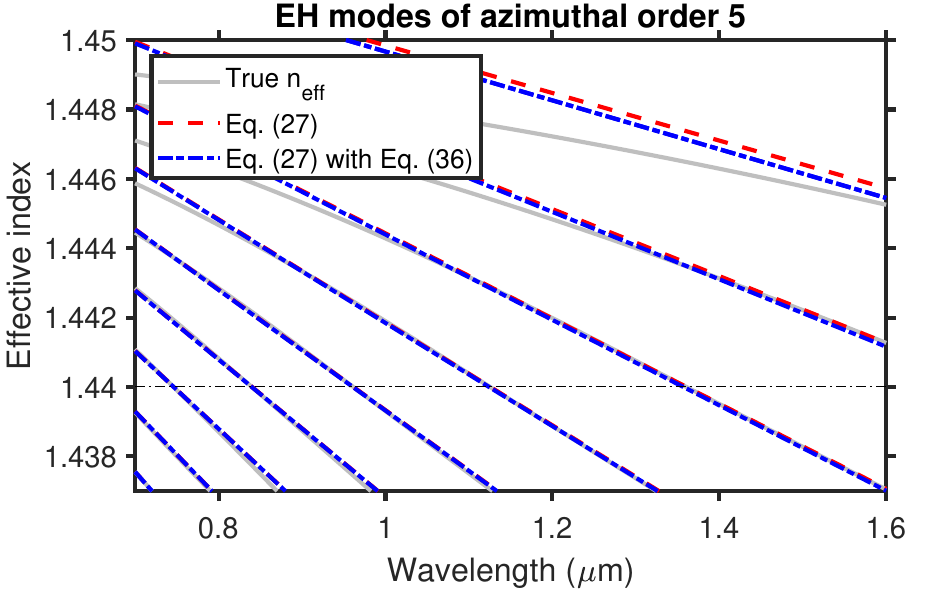}
\caption{The effective index of EH-modes with $m = 5$ in the example fiber. The solid gray lines show the true effective indices. The dashed red lines show the approximate effective index using  Eq.~(\ref{eq:HEneff}), and the dot-dash blue line shows the cascaded approximation with Eq.~(\ref{eq:Jmzeroappr}) plugged into Eq.~(\ref{eq:HEneff}). Note that even the cascaded approximation works so well that the red dashed lines are not even visible for higher radial mode orders.}
\label{fig3}
\end{figure}

To see how well the approximation in Eq.~(\ref{eq:Jmzeroappr}) works with that of Eq.~(\ref{eq:EHneff}), consider EH-modes of azimuthal order $m = 5$ in the same example fiber. The lower azimuthal order allows for more radial orders to be supported in the fiber. Figure~\ref{fig3} shows the true effective indices and approximation of Eq.~(\ref{eq:EHneff}) with and without the approximation of Eq.~(\ref{eq:Jmzeroappr}). The cascaded approximation quickly becomes indistinguishable from the true linear approximation as the radial mode order increases. This means that Eq.~(\ref{eq:EHneff}) together with Eq.~(\ref{eq:Jmzeroappr}) yield a very reasonable approximation for the effective indices of EH modes or large radial order without even having to determine or know the Bessel function zeros (mode cutoff wavelengths). Remarkably, one only needs to plug in the fiber parameters and azimuthal and radial mode order to use the approximation.

\section{Conclusions, Implications, and Generalisations}

Linear approximations for mode effective indices near the cutoff wavelengths were derived through Taylor and Puiseux series of Bessel functions. The approximations are presented in Eqs.~(\ref{eq:HEneff}) and (\ref{eq:EHneff}), and they have profound implications:

\begin{enumerate}
  \item They are simple. They avoid the usual trial-and-error search for the effective indices, and simply give the effective index as a function of wavelength, fiber parameters, and azimuthal mode order. 
  \item They allowed for the derivation of certain surprising analytical results for the first time, such as the EH-mode group velocity at cutoff only depending on wavelength through material dispersion, and otherwise being independent of (the cutoff) wavelength, fiber core radius, or even radial mode order. 
  \item They were used to derive previously unknown asymptotic properties of modes near cutoff as their azimuthal and/or radial mode order is increased.
  \item They can be utilized in speeding up effective index calculations, as they provide an educated guess for the effective index in the vicinity of the cutoff wavelength, and elsewhere through extrapolation.
  \item They remain valid beyond cutoff, where the effective indices cannot even be determined in the usual manner, and they give the real part of the effective index of lossy modes.
  \item They exposed (and avoid) the numerical problems with determining the effective indices near cutoff the traditional way by matching the tangential fields, which results in having to subtract two very large numbers from another to yield something very small.
\end{enumerate}

The method derived here also allows for some obvious generalisations. The series expansion for the determinant in the modal determinant equation was terminated after the first two dominant terms, yielding the linear approximation. Naturally, more terms could be kept in the series expansion, making it possible to analytically derive expressions for the group velocity dispersion in the vicinity of cutoff wavelengths from the quadratic approximation, for example. Furthermore, even though this manuscript only considers a simple step-index fiber consisting of a circular core and infinite cladding, the same technique can be applied to derive analytical results for more complicated fiber designs, as long as the fiber is circularly symmetric, the effective index is piecewise constant, and the fiber has cutoff conditions related to the zeros of a function (such as the Bessel function $J_m$ here).

The circularity requirement excludes, for example, hexagonal photonic crystal fibers, but the circularity is not a strict requirement in the sense that the same techniques and ideas could be applied to rectangular waveguides and others. The piecewise-constant-$n$-condition excludes graded-index fibers, and the cutoff requirement excludes Bragg fibers where the guiding mechanism is not total (internal) reflection. The method is, however, readily generalizable to ring-core fibers, dispersion-compensating fibers, metal-clad hollow-core fibers, etc. Determining the modal effective indices for all such fibers always leads to the determinant of a matrix involving Bessel functions $J_m$, $Y_m$, and $K_m$ having to equal zero, and this determinant can be expanded as a series like was done here. As a final note, since Bessel functions are associated with a lot of other physical phenomena, such as the vibrations of a drum head, the mathematical technique introduced here might offer insight and benefits in other branches of physics as well.

\pagebreak
\onecolumn
\begin{center}
\textbf{\titlefont Analytical Expressions for Effective Indices of Modes of Optical Fibers Near and Beyond Cutoff - Supplemental Document}
\end{center}

\setcounter{equation}{0}
\setcounter{section}{0}
\setcounter{figure}{0}
\setcounter{table}{0}
\setcounter{page}{1}
\makeatletter
\renewcommand{\theequation}{S\arabic{equation}}
\renewcommand{\thefigure}{S\arabic{figure}}
\renewcommand{\bibnumfmt}[1]{[S#1]}
\renewcommand{\citenumfont}[1]{S#1}

\section{Derivation of the First-Order Determinant Equation}

\subsection{The Determinant without Approximations}

Consider a dielectric step-index fiber with a circular core of radius $a$ and refractive index $n_1$ and an infinite cladding of refractive index $n_2$. The electric and magnetic field $z$-components of a mode in such a fiber can be written in polar coordinates $r,z$ as
\begin{align}
    E_z(r,z,t) &= A J_m(pr) e^{i m \phi} e^{i(\beta z - \omega t)} \\
    H_z(r,z,t) &= B J_m(pr) e^{i m \phi} e^{i(\beta z - \omega t)} 
\end{align}
in the core and 
\begin{align}
    E_z(r,z,t) &= C K_m(qr) e^{i m \phi} e^{i(\beta z - \omega t)} \\
    H_z(r,z,t) &= D K_m(qr) e^{i m \phi} e^{i(\beta z - \omega t)} 
\end{align}
in the cladding,
where $A, B, C, D$ are constants, $J_m$ is the Bessel function of the first kind, $K_m$ is the modified Bessel function of the second kind, $p = k_0 \sqrt{n_1^2 - \neff^2}$, $q = k_0 \sqrt{\neff^2 - n_2^2}$, and $k_0 = 2 \pi / \lambda_0$ is the vacuum wave number. The propagation constant $\beta$ is related to the effective index $\neff$ through $\beta = 2 \pi \neff / \lambda_0$. The transverse components can be gotten from $E_z$ and $H_z$ through
\begin{align}
	E_r
	&= 
		\frac{i \beta}{k_c^2} \left[ \pd_r E_z
		+
		\frac{\omega \mu}{\beta}  \frac{1}{r} \pd_\varphi H_z \right]
	 \\
	H_r
	&=
	\frac{i \beta}{k_c^2} \left[ - \frac{\omega \varepsilon}{\beta} \frac{1}{r} \pd_\varphi E_z + \pd_r H_z \right]  
\end{align}
and
\begin{align}
	E_\phi
	&= 
		\frac{i \beta}{k_c^2} \left[\frac{1}{r} \pd_\varphi  E_z
		-
		\frac{\omega \mu}{\beta} \pd_r H_z \right]
	 \\
	H_\phi
	&=
	\frac{i \beta}{k_c^2} \left[ \frac{\omega \varepsilon}{\beta} \pd_r E_z + \frac{1}{r} \pd_\varphi H_z \right]  ,
\end{align}
where $k_c^2 = p^2$ in the core and $k_c^2 = - q^2$ in the cladding, and $\varepsilon$ and $\mu$ are the permittivity and permeability of the medium ($\varepsilon$ is different in the core vs. the cladding). In the core, we have
\begin{align}
    	E_\phi
	&= 
		\frac{i \beta}{p^2} 
  \left[
            A \frac{i m}{r} J_m(pr) 
		-
		      B \frac{\omega \mu_1}{\beta} p \frac{J_{m-1} (pr) - J_{m+1} (pr)}{2}  
 \right] e^{i m \phi} e^{i(\beta z - \omega t)}
	 \\
	H_\phi
	&=
	\frac{i \beta}{p^2} \left[ 
 A \frac{\omega \varepsilon_1}{\beta} p \frac{J_{m-1} (pr) - J_{m+1} (pr)}{2}
 + 
 B \frac{i m}{r} J_m(pr)  \right] e^{i m \phi} e^{i(\beta z - \omega t)} ,
\end{align}
and in the cladding the corresponding expressions are 
\begin{align}
    	E_\phi
	&= 
		\frac{i \beta}{-q^2} 
  \left[
            C \frac{i m}{r} K_m(qr) 
		+
		      D \frac{\omega \mu_2}{\beta} q \frac{K_{m-1} (qr) + K_{m+1} (qr)}{2}  
 \right] e^{i m \phi} e^{i(\beta z - \omega t)}
	 \\
	H_\phi
	&=
	\frac{i \beta}{-q^2} \left[ 
 - C \frac{\omega \varepsilon_2}{\beta} q \frac{K_{m-1} (qr) + K_{m+1} (qr)}{2}
 + 
 D \frac{i m}{r} K_m(qr)  \right] e^{i m \phi} e^{i(\beta z - \omega t)} .
\end{align}
The tangential components of the electric and magnetic fields ($E_z, E_\phi, H_z, H_\phi$) need to be continuous across the core-cladding interface (at $r=a$). In matrix form, these continuity conditions can be written as
\begin{equation}
    M \begin{bmatrix}
        A \\ B \\ C \\ D
    \end{bmatrix}
    =
    \begin{bmatrix}
    J_m & 0 & -K_m & 0 \\
    0 & J_m & 0 & -K_m  \\
 i \frac{m \beta}{a p^2} J_m  
 & 		
 - \frac{\omega \mu_1}{p} J_m'  
 & 
i \frac{m \beta}{a q^2} K_m
 & 
- \frac{\omega \mu_2}{q} K_m' 
\\
\frac{\omega \varepsilon_1}{p} J_m' 
&
i \frac{m \beta}{a p^2} J_m
& 
\frac{\omega \varepsilon_2}{q} K_m'
&
i \frac{m \beta}{a q^2} K_m
    \end{bmatrix}
    \begin{bmatrix}
        A \\ B \\ C \\ D
    \end{bmatrix}
    =
        \begin{bmatrix}
        0 \\ 0 \\ 0 \\ 0 
    \end{bmatrix} ,
\end{equation}
where $J_m = J_m(pa)$, $K_m = K_m(qa)$, $J_m' = [J_{m-1} (pa) - J_{m+1} (pa)]/2$, $K_m' = -[K_{m-1} (qa) + K_{m+1} (qa)]/2$, and the last two rows have been multiplied by $-i$. The matrix equation has a solution if and only if the determinant of the matrix $M$ is zero.
\begin{align}
   \det(M) = 
    &J_m \left\{ J_m \left[  - \left(  \frac{m \beta}{a q^2} K_m \right)^2
    + \mu_2 \varepsilon_2 \left(\frac{\omega}{q} K_m' \right)^2 \right] \right\} \\
    &+
    J_m \left\{ - K_m  \left[  - \mu_1 \varepsilon_2 \frac{\omega^2}{p q} J_m' K_m'
                              + J_m K_m \left( \frac{m \beta}{a p q} \right)^2 \right] \right\} 
                             \nonumber  \\
&- 
    K_m \left\{  - J_m \left[  - \left( \frac{m \beta}{a p q} \right)^2 J_m K_m + \mu_2 \varepsilon_1 \frac{\omega^2}{p q} J_m' K_m' \right]  \right\} \nonumber  \\
&- 
    K_m \left\{  - K_m \left[  - \left( \frac{m \beta}{a p^2} J_m \right)^2 
                               + \mu_1 \varepsilon_1 \left( \frac{\omega}{p} J_m' \right)^2 \right]  \right\} \nonumber  \\
            = 
    &  \underbrace{- \left[  \frac{m \beta}{a} J_m K_m \left( \frac{1}{p^2} + \frac{1}{q^2} \right)  \right]^2}_{\stackrel{\text{def}}{=} T_1}
       + \underbrace{\mu_2 \varepsilon_2 \left(\frac{\omega}{q} J_m K_m' \right)^2}_{\stackrel{\text{def}}{=} T_2} 
       + \underbrace{\mu_1 \varepsilon_1 \left(\frac{\omega}{p} K_m J_m' \right)^2}_{\stackrel{\text{def}}{=} T_3}  
			 + \underbrace{(\mu_1 \varepsilon_2 + \mu_2 \varepsilon_1) \frac{\omega^2}{p q} J_m K_m J_m' K_m'}_{\stackrel{\text{def}}{=} T_4}   . \label{suppl_eq:fulldet}
\end{align}
The determinant thus consist of a sum of four terms, labeled $T_1, T_2, T_3$, and $T_4$, that are products of various Bessel functions and other factors.

\subsection{Series for Bessel Functions Near Cutoff}

Near cutoff $\neff \approx n_2$, so we can write $\neff = n_2 + d$, where $d$ is small. The idea behind the approximation is to express the determinant as a power series (Puiseux series) in $d$ and only keep the terms of the two lowest orders in $d$. To write the determinant as a Puiseux series of $d$ we first note that the Bessel function terms $J_m$ and $J_m'$ have Taylor expansions around $d = 0$, so they can simply be written as such for the full expansion. Let us denote these Taylor expansions as
\begin{equation}
    J_m(pa) = A_m + B_m d + \mathcal{O}(d^2) ,
\end{equation}
where
\begin{align}
    A_m &= J_m(V) \label{suppl_eq:Am} \\
    B_m &=  -  \frac{a^2 k_0^2 n_2}{2 V} [J_{m-1}(V) - J_{m+1}(V)] 
         =  -  \frac{a^2 k_0^2 n_2}{2 V} [A_{m-1} - A_{m+1}]   . \label{suppl_eq:Bm}
\end{align}
The parameter $q$ becomes zero at $d = 0$, and the modified Bessel functions diverge at the origin, so they cannot be expanded as Taylor series. They do, however, admit to more complicated series expansions (Eq. 9.6.11 on page 375 of \cite{book:stegun}) of the form
\begin{align}
    K_m = K_m(qa) = \frac{1}{(a q)^m} \left[ D_m + F_m (a q)^2 + G_m (a q)^4 + \mathcal{O}(d^3 \ln d) \right] ,
\end{align}
where $D_m, F_m$, and $G_m$ are of the functional form $u_m + v_m \ln(q a)$ with $u_m$ and $v_m$ constants. The series expansion inside the brackets only contains even powers of $aq$, and hence the error term inside is $\mathcal{O}[ (a q)^6 \ln(aq) ] = \mathcal{O}[d^3 \ln(d) ]$. The series expansion coefficients are
\begin{align}
    D_0 &= \ln 2 - \gamma - \ln (aq) \label{suppl_eq:D0}  \\ 
    F_0 &= \frac{1}{4} \left[  1 - \gamma + \ln 2 - ln (a q)  \right] \label{suppl_eq:F0}  \\ 
    G_0 &= \frac{1}{64} \left[ \frac{3}{2} - \gamma + \ln 2 - \ln (a q)  \right] ,  \label{suppl_eq:G0}  \
\end{align}

\begin{align}
    D_1 &= 1 = 2^{1 - 1} (1 - 1)!     \label{suppl_eq:D1}  \\ 
    F_1 &= \frac{1}{4} \left[ 2 \gamma - 1 - 2 \ln 2 + 2 \ln (a q)  \right] \label{suppl_eq:F1}  \\ 
    G_1 &= \frac{1}{16} \left[ \gamma - \frac{5}{4} - \ln 2 + \ln (a q)  \right] , \label{suppl_eq:G1}  
\end{align}

\begin{align}
    D_2 &= 2 = 2^{2 - 1} (2 - 1)!   \label{suppl_eq:D2}  \\ 
    F_2 &= - \frac{1}{2} = - 2^{2 - 3} (2 - 2)!  \label{suppl_eq:F2}  \\ 
    G_2 &= \frac{1}{8} \left[ \frac{3}{4} - \gamma + \ln 2 - \ln (a q)  \right] ,  \label{suppl_eq:G2}  
\end{align}
and, finally, for $m \geq 3$,
\begin{align}
    D_m &=   2^{m - 1} (m - 1)!  \label{suppl_eq:Dm}  \\ 
    F_m &= - 2^{m - 3} (m - 2)!  \label{suppl_eq:Fm}  \\ 
    G_m &=   2^{m - 6} (m - 3)!  \label{suppl_eq:Gm}  .
\end{align}
We can see readily some important properties:
\begin{align}
    D_{m + 1} &= 2 m D_m ~~~\text{ for }~ m \geq 1 , \label{suppl_eq:DmpDm} \\
    F_{m + 1} &= 2 (m - 1) F_m ~~~\text{ for }~ m \geq 2 , \label{suppl_eq:FmpFm} \\
    G_{m + 1} &= 2 (m - 2) G_m ~~~\text{ for }~ m \geq 3 . \label{suppl_eq:GmpGm}
\end{align}
Furthermore, for $m \geq 3$, $D_m, F_m$, and $G_m$ are constants, and for $m \geq 4$ the error term also reduces to $ \mathcal{O}(d^3)$.

\subsection{Series Expansion of the Determinant}

We first note that both $T_1$ and $T_2$ have a factor $J_m^2$, so we can account for this factor later and expand the rest first. Keeping the three dominant orders for $T_1 / J_m^2$, we get
\begin{align}
    \frac{T_1}{J_m^2} =  &- \frac{m^2 k_0^2 (n_2 + d)^2 }{a^2}  K_m^2 \left( \frac{1}{p^2} + \frac{1}{q^2} \right)^2 \\
    \approx
    &- \frac{m^2 k_0^2 (n_2^2 + 2 n_2 d + d^2) }{a^2 (a q)^{2m}}  [ D_m + F_m (a q)^2 + G_m (a q)^4 ]^2 \left( \frac{1}{p^4} + \frac{2}{p^2 q^2} + \frac{1}{q^4} \right) \\
    \approx
    &- \frac{m^2 k_0^2 (n_2^2 + 2 n_2 d + d^2) }{a^2 (a q)^{2m}}  [ D_m^2 + 2 D_m F_m (a q)^2 + ( F_m^2 + 2 D_m G_m ) (a q)^4 ] \left( \frac{1}{p^4} + \frac{2}{p^2 q^2} + \frac{1}{q^4} \right) 
\end{align}
\begin{align}
    \frac{T_1}{J_m^2}
    \approx
    &- \frac{m^2 k_0^2 (n_2^2 + 2 n_2 d + d^2) }{a^2 (a q)^{2m}}  [ D_m^2 + 2 D_m F_m (a q)^2 + ( F_m^2 + 2 D_m G_m ) (a q)^4 ] \frac{1}{p^4}  \\
    &- \frac{m^2 k_0^2 (n_2^2 + 2 n_2 d + d^2) }{a^2 (a q)^{2m}}  [ D_m^2 + 2 D_m F_m (a q)^2 + ( F_m^2 + 2 D_m G_m ) (a q)^4 ] \frac{2}{p^2 q^2} \nonumber \\
    &- \frac{m^2 k_0^2 (n_2^2 + 2 n_2 d + d^2) }{a^2 (a q)^{2m}}  [ D_m^2 + 2 D_m F_m (a q)^2 + ( F_m^2 + 2 D_m G_m ) (a q)^4 ] \frac{1}{q^4} \nonumber
    \end{align}
\begin{align}
    \frac{T_1}{J_m^2}
        \approx
    &- \frac{m^2 k_0^2 n_2^2 }{a^2 (a q)^{2m}} D_m^2 \frac{1}{p^4}  \\
    &- \frac{m^2 k_0^2 (n_2^2 + 2 n_2 d) }{a^2 (a q)^{2m}}  [ D_m^2 + 2 D_m F_m (a q)^2 ] \frac{2}{p^2 q^2} \nonumber \\
    &- \frac{m^2 k_0^2 (n_2^2 + 2 n_2 d + d^2) }{a^2 (a q)^{2m}}  [ D_m^2 + 2 D_m F_m (a q)^2 + ( F_m^2 + 2 D_m G_m ) (a q)^4 ] \frac{1}{q^4} \nonumber
\end{align}

\begin{align}
    \frac{T_1}{J_m^2}
        \approx
    &- \frac{m^2 k_0^2}{a^2 (a q)^{2m}} n_2^2 D_m^2 \frac{1}{p^4}  \\
    &- \frac{m^2 k_0^2}{a^2 (a q)^{2m}}  [ n_2^2 D_m^2 + 2 n_2 D_m^2 d + 2 n_2^2 D_m F_m (a q)^2 ] \frac{2}{p^2 q^2} \nonumber \\
    &- \frac{m^2 k_0^2}{a^2 (a q)^{2m}}  [ 
        n_2^2 D_m^2 + 2 n_2 D_m^2 d + D_m^2 d^2 
    + 2 D_m F_m (a q)^2 (n_2^2 + 2 n_2 d) ] \frac{1}{q^4} \nonumber \\
    &- \frac{m^2 k_0^2}{a^2 (a q)^{2m}}  n_2^2 ( F_m^2 + 2 D_m G_m ) (a q)^4  \frac{1}{q^4} \nonumber
\end{align}

Re-organising some terms and introducing certain factors for reasons that will become clear shortly yields
\begin{align}
    \frac{T_1}{J_m^2}
        \approx
    &- \frac{k_0^2}{4 (a q)^{2m + 4}} 4 m^2 n_2^2 D_m^2 \frac{(a q)^4}{a^2 p^4}  \\
    &- \frac{k_0^2}{4 (a q)^{2m + 4}} 4 m^2 [ n_2^2 D_m^2 (a q)^2   +   2 n_2 D_m^2 d (a q)^2  +   2 n_2^2 D_m F_m (a q)^4 ] \frac{2}{p^2} \nonumber \\
    &- \frac{k_0^2}{4 (a q)^{2m + 4}} 4 m^2 a^2 [  n_2^2 D_m^2 + 2 n_2 D_m^2 d + D_m^2 d^2 + 2 D_m F_m (a q)^2 (n_2^2 + 2 n_2 d) ] \nonumber   \\
    &- \frac{k_0^2}{4 (a q)^{2m + 4}} 4 m^2 a^2  n_2^2 ( F_m^2 + 2 D_m G_m ) (a q)^4  \nonumber
\end{align}

Note that further approximations can be made due to the $q^{2 m}$ factors in the denominator, but we will leave these as they are for now. Similarly, for the second term $T_2$ we have
\begin{align}
    \frac{T_2}{J_m^2} =  
    &  \mu_2 \varepsilon_2 \left(\frac{\omega}{q} K_m' \right)^2 
   = \frac{\mu_2 \varepsilon_2 \omega^2}{q^2}   \left[ -  \frac{ K_{m-1} + K_{m+1} }{2} \right]^2 \\
   \approx &\frac{\mu_2 \varepsilon_2 \omega^2}{4 q^2}  
   \left[ \frac{D_{m-1} + F_{m-1} (a q)^2 + G_{m-1} (a q)^4}{(a q)^{m-1}}
        + \frac{D_{m+1} + F_{m+1} (a q)^2 + G_{m+1} (a q)^4}{(a q)^{m+1}}  \right]^2 \\
  \approx &\frac{n_2^2 k_0^2}{4 q^2 (a q)^{2m + 2}}  
   \left[ D_{m-1} (a q)^2 + F_{m-1} (a q)^4  
        + D_{m+1} + F_{m+1} (a q)^2 + G_{m+1} (a q)^4   \right]^2 ,
\end{align}
where we have used $\mu_2 \varepsilon_2 \omega^2 = n_2^2 k_0^2$. Expanding the bracketed square and neglecting terms that are proportional to $d^3$ or higher powers yields
\begin{align}
    \frac{T_2}{J_m^2} 
  \approx &\frac{k_0^2}{4 (a q)^{2m + 4}} a^2 n_2^2  D_{m+1}^2  \\
     + & 2 \frac{k_0^2}{4 (a q)^{2m + 4}}  a^2 n_2^2 D_{m+1} (D_{m-1} + F_{m+1}) (a q)^2  \nonumber \\
        + &\frac{k_0^2}{4 (a q)^{2m + 4}}  a^2 n_2^2 \left[2 D_{m+1} F_{m-1} + 2 D_{m+1} G_{m+1} + (D_{m-1} + F_{m+1})^2 \right] (a q)^4  \nonumber
\end{align}
Since $D_{m + 1} = 2 m D_m$ for $m \geq 1$, the dominant term in $T_2$ is exactly minus the dominant term in $T_1$, and they cancel each other out when $T_1$ and $T_2$ are added together to get the full determinant. This is why we needed to keep the second-order terms $d^2$ to get an eventual first-order approximation for the determinant. The dominant order in the sum $T_1 + T_2$ is thus $1/(a q)^{2 m + 2} \propto 1/d^{m + 1}$, and the next order is $1/d^m$. The sum $(T_1 + T_2)/J_m^2$ is
\begin{align}
    \frac{4 (a q)^{2m + 4}}{k_0^2} \frac{T_1 + T_2}{J_m^2} 
  \approx 
       &    2 a^2 n_2^2 D_{m+1} (D_{m-1} + F_{m+1}) (a q)^2  \\
       &+     a^2 n_2^2 \left[2 D_{m+1} F_{m-1} + 2 D_{m+1} G_{m+1} + (D_{m-1} + F_{m+1})^2 \right] (a q)^4 \nonumber \\
       &-   4 m^2 n_2^2 D_m^2 \frac{(a q)^4}{a^2 p^4} \nonumber \\
       &-   4 m^2 [ n_2^2 D_m^2 (a q)^2   +   2 n_2 D_m^2 d (a q)^2  +   2 n_2^2 D_m F_m (a q)^4 ] \frac{2}{p^2} \nonumber \\
       &-   4 m^2 a^2 [ D_m^2 (a q)^2 / (a k_0)^2+ 2 D_m F_m (a q)^2 (n_2^2 + 2 n_2 d) ] \nonumber \\
       &-   4 m^2 a^2  n_2^2 ( F_m^2 + 2 D_m G_m ) (a q)^4 , \nonumber
\end{align}
where the dominant terms have cancelled each other out and we have used $2 n_2 d + d^2 = (a q)^2 / (a k_0)^2$. Now every term on both sides has a factor $(a q)^2$ that can be divided out:
\begin{align}
    \frac{4 (a q)^{2m + 2}}{k_0^2} \frac{T_1 + T_2}{J_m^2} 
  \approx 
       &    2 a^2 n_2^2 D_{m+1} (D_{m-1} + F_{m+1})  \\
       &+     a^2 n_2^2 \left[2 D_{m+1} F_{m-1} + 2 D_{m+1} G_{m+1} + (D_{m-1} + F_{m+1})^2 \right] (a q)^2 \nonumber \\
       &-   4 m^2 n_2^2 D_m^2 \frac{(a q)^2}{a^2 p^4} \nonumber \\
       &-   4 m^2 [ n_2^2 D_m^2 +   2 n_2 D_m^2 d +   2 n_2^2 D_m F_m (a q)^2 ] \frac{2}{p^2} \nonumber \\
       &-   4 m^2 a^2 [ D_m^2 / (a k_0)^2 + 2 D_m F_m (n_2^2 + 2 n_2 d) ] \nonumber \\
       &-   4 m^2 a^2  n_2^2 ( F_m^2 + 2 D_m G_m ) (a q)^2 . \nonumber
\end{align}
On the right-hand side, the dominant terms are now constants, so we only need to keep these and the terms proportional to $d$. This means that we can approximate the factors $(a q)^2$ as $(a q)^2 = a^2 k_0^2 (2 n_2 d + d^2) \approx 2 a^2 k_0^2 n_2 d$. This yields
\begin{align}
    \frac{2 (a q)^{2m + 2}}{a^2 k_0^2} \frac{T_1 + T_2}{J_m^2} 
  \approx 
       &      n_2^2 D_{m+1} (D_{m-1} + F_{m+1})  \\
       &+     a^2 k_0^2 n_2^3 \left[2 D_{m+1} F_{m-1} + 2 D_{m+1} G_{m+1} + (D_{m-1} + F_{m+1})^2 \right] d \nonumber \\
       &-   4 a^2 k_0^2 m^2 n_2^3 D_m^2 \frac{1}{(a p)^4} d \nonumber \\
       &-   4 m^2 [ n_2^2 D_m^2 +   2 n_2 D_m^2 d +   4 a^2 k_0^2 n_2^3 D_m F_m d  ] \frac{1}{(a p)^2} \nonumber \\
       &-   2 m^2 [ D_m^2 / (a k_0)^2 + 2 D_m F_m (n_2^2 + 2 n_2 d) ] \nonumber \\
       &-   4 m^2 n_2^3 ( F_m^2 + 2 D_m G_m ) a^2 k_0^2 d , \nonumber
\end{align}
where both sides have been divided by 2. The factor $p$ still contains $d$, and terms involving $p$ thus need to be expanded to first order. We have
\begin{equation}
    a p \approx V \left(1 - \frac{n_2 d}{n_1^2 - n_2^2} \right)  
    =  V \left(1 - \frac{a^2 k_0^2 n_2 d}{V^2} \right) ,
\end{equation}
and
\begin{equation}
    \frac{1}{a p} \approx \frac{1}{V} \left(1 + \frac{n_2 d}{n_1^2 - n_2^2} \right) 
    = \frac{1}{V} \left(1 + \frac{a^2 k_0^2 n_2 d}{V^2} \right),
\end{equation}
which simply follow from $\sqrt{1 + x} \approx 1 + x/2$ and $1 / (1 - x) \approx 1 + x$ for small $x$. Therefore,
\begin{align}
    \frac{2 (a q)^{2m + 2}}{a^2 k_0^2} \frac{T_1 + T_2}{J_m^2} 
  \approx 
       &      n_2^2 D_{m+1} (D_{m-1} + F_{m+1})  \\
       &+     a^2 k_0^2 n_2^3 \left[2 D_{m+1} F_{m-1} + 2 D_{m+1} G_{m+1} + (D_{m-1} + F_{m+1})^2 \right] d \nonumber \\
       &-   4 a^2 k_0^2 m^2 n_2^3 D_m^2 \frac{1}{V^4} d \nonumber \\
       &-   4 m^2 [ n_2^2 D_m^2 +   2 n_2 D_m^2 d +   4 a^2 k_0^2 n_2^3 D_m F_m d  ]  \frac{1}{V^2}  \left(  1 + 2 \frac{a^2 k_0^2 n_2 d}{V^2}   \right)  \nonumber \\
       &-   2 m^2 [ D_m^2 / (a k_0)^2 + 2 D_m F_m (n_2^2 + 2 n_2 d) ] \nonumber \\
       &-   4 m^2 n_2^3 ( F_m^2 + 2 D_m G_m ) a^2 k_0^2 d , \nonumber
\end{align}
Rearranging to separate the orders in $d$ gives
\begin{align}
    \frac{2 (a q)^{2m + 2}}{a^2 k_0^2} \frac{T_1 + T_2}{J_m^2} 
  \approx 
       &      n_2^2 D_{m+1} (D_{m-1} + F_{m+1})    -   4 m^2 n_2^2 D_m^2 \frac{1}{V^2}    \\
       &-  2 m^2 D_m^2 / (a k_0)^2 - 4 m^2 n_2^2 D_m F_m \nonumber \\
       &+     a^2 k_0^2 n_2^3 \left[2 D_{m+1} F_{m-1} + 2 D_{m+1} G_{m+1} + (D_{m-1} + F_{m+1})^2 \right] d \nonumber \\
       &-   4 a^2 k_0^2 m^2 n_2^3 \left[ 3 \frac{D_m^2}{V^4}   +   F_m^2 + 2 D_m G_m \right]   d \nonumber \\ 
       &-    8 m^2 n_2  D_m \left[ \frac{D_m }{V^2}  +  2 \frac{ a^2 k_0^2 n_2^2 F_m }{V^2} + F_m \right] d . \nonumber
\end{align}
The remaining thing to do is to multiply both sides by $J_m^2$ and neglect all but the two dominant orders. Since the orders have already been separated in the expression above, the easiest way is to expand $J_m^2$ first:
\begin{equation}
    J_m^2 \approx (A_m + B_m d)^2 \approx A_m^2 + 2 A_m B_m d ,
\end{equation}
and multiply both sides by this:
\begin{align}
    \frac{2 (a q)^{2m + 2}}{a^2 k_0^2} (T_1 + T_2) 
  \approx 
       &      n_2^2 A_m^2 D_{m+1} (D_{m-1} + F_{m+1})  
         -   4 m^2 n_2^2 A_m^2 D_m^2 \frac{1}{V^2} \\
       & -    2 m^2 A_m^2 D_m^2 / (a k_0)^2 
         - 4 m^2 n_2^2 A_m^2 D_m F_m \nonumber \\
       &+     a^2 k_0^2 n_2^3 A_m^2 \left[2 D_{m+1} F_{m-1} + 2 D_{m+1} G_{m+1} + (D_{m-1} + F_{m+1})^2 \right] d \nonumber \\
       &-   4 a^2 k_0^2 m^2 n_2^3 A_m^2 \left[ 3 \frac{D_m^2}{V^4}   +   F_m^2 + 2 D_m G_m \right]   d \nonumber \\ 
       &-    8 m^2 n_2 A_m^2 D_m \left[ \frac{D_m }{V^2}  +  2 \frac{ a^2 k_0^2 n_2^2 F_m }{V^2} + F_m \right] d \nonumber \\
       & +   2 n_2^2 A_m B_m D_{m+1} (D_{m-1} + F_{m+1})  d
         -   8 m^2 n_2^2 A_m B_m D_m^2 \frac{1}{V^2} d \nonumber \\
       & -    4 \frac{m^2}{(a k_0)^2} A_m B_m D_m^2 d
         - 8 m^2 n_2^2 A_m B_m D_m F_m d \nonumber
\end{align}

The approximation for the third term $T_3$ is
\begin{align}
    T_3 &= \mu_1 \varepsilon_1 \left(\frac{\omega}{p} K_m J_m' \right)^2 \\
        &\approx \frac{n_1^2 k_0^2}{p^2} J_m'^2 \frac{1}{(a q)^{2 m}} \left(  D_m + F_m (a q)^2 + G_m (a q)^4 \right)^2 .
\end{align}
We can readily see that the lowest order in the approximation for $T_3$ is $1/d^m$, and since the sum $T_1 + T_2$ contains terms of order $1/d^{m+1}$, we only need to keep the lowest order term in the approximation for $T_3$. Thus:
\begin{align}
    T_3 &\approx \frac{n_1^2 k_0^2}{p^2} \left(  \frac{J_{m-1} - J_{m+1}}{2}   \right)^2 \frac{1}{(a q)^{2 m}} D_m^2 \\
        &\approx \frac{n_1^2 k_0^2}{p^2} (  A_{m-1} - A_{m+1}   )^2 \frac{D_m^2}{4 (a q)^{2 m}} \\
        &\approx \frac{n_1^2 k_0^2}{p^2} (  A_{m-1}^2 - 2 A_{m-1} A_{m+1} + A_{m+1}^2  ) \frac{D_m^2}{4 (a q)^{2 m}} ,
\end{align}
and hence
\begin{align}
    \frac{2 (a q)^{2 m + 2}}{a^2 k_0^2} T_3
        &\approx \frac{(a k_0)^2 n_1^2}{(a p)^2} (  A_{m-1}^2 - 2 A_{m-1} A_{m+1} + A_{m+1}^2  ) D_m^2 \frac{(a q)^2}{2}   \\
        &\approx \frac{(a k_0)^2 n_1^2 n_2}{(a p)^2} (  A_{m-1}^2 - 2 A_{m-1} A_{m+1} + A_{m+1}^2  ) D_m^2 d \\
        &\approx \frac{(a k_0)^2 n_1^2 n_2}{V^2} (  A_{m-1}^2 - 2 A_{m-1} A_{m+1} + A_{m+1}^2  ) D_m^2 d 
\end{align}

The approximation for $T_4$, when keeping only terms of orders $1/d^{m+1}$ and $1/d^m$, is
\begin{align}
    T_4 &=  (\mu_1 \varepsilon_2 + \mu_2 \varepsilon_1) \frac{\omega^2}{p q} J_m K_m J_m' K_m' \\
        &\approx - (n_1^2 + n_2^2) \frac{k_0^2}{p q} J_m J_m' 
        \frac{1}{(a q)^m} \left[ D_m + F_m (a q)^2 \right]  
   \left[  \frac{D_{m-1}}{2 (a q)^{m-1}}
         + \frac{D_{m+1} + F_{m+1} (a q)^2}{2 (a q)^{m+1}} \right] \\
         &\approx - (n_1^2 + n_2^2) \frac{a k_0^2 J_m J_m'}{2 p (a q)^{2m+2}}  
        \left[ D_m + F_m (a q)^2 \right]  
   \left[  D_{m-1} (a q)^2 + D_{m+1} + F_{m+1} (a q)^2 \right] \\
   &\approx - (n_1^2 + n_2^2) \frac{a k_0^2 J_m J_m'}{2 p (a q)^{2m+2}}  
   \left[  D_m D_{m+1}
         + (D_m D_{m-1} 
         + D_m F_{m+1}  
         + D_{m+1} F_m )(a q)^2
         \right] \\
   &\approx - (n_1^2 + n_2^2) \frac{a k_0^2 J_m J_m'}{2 p (a q)^{2m+2}}  
   \left[  D_m D_{m+1}
         + 2 a^2 k_0^2 n_2 (D_m D_{m-1} 
         + D_m F_{m+1}  
         + D_{m+1} F_m ) d
         \right] .
\end{align}
Expanding $J_m J_m'$:
\begin{align}
    J_m J_m' &\approx \left( A_m + B_m d \right) \left[ A_{m-1} - A_{m+1}  + (B_{m-1} - B_{m+1}) d \right] / 2 \\
    &\approx   \frac{A_m}{2} (A_{m-1} - A_{m+1})  + \frac{A_m}{2} (B_{m-1} - B_{m+1}) d  + \frac{B_m}{2} ( A_{m-1} - A_{m+1} ) d .
\end{align}
Thus,
\begin{align}
    \frac{2 (a q)^{2m+2}}{a^2 k_0^2} T_4 
    &\approx - \frac{n_1^2 + n_2^2}{2 a p} A_m (A_{m-1} - A_{m+1})  D_m D_{m+1} \\
    &- \frac{n_1^2 + n_2^2}{a p} a^2 k_0^2 n_2  A_m (A_{m-1} - A_{m+1}) (D_m D_{m-1}  + D_m F_{m+1}  + D_{m+1} F_m ) d \nonumber \\
       &- \frac{n_1^2 + n_2^2}{2 a p} \left[ A_m (B_{m-1} - B_{m+1})  + B_m ( A_{m-1} - A_{m+1} ) \right] D_m D_{m+1} d \nonumber \\
&\approx - \frac{n_1^2 + n_2^2}{2 V} A_m (A_{m-1} - A_{m+1})  D_m D_{m+1} \\
        &- \frac{n_1^2 + n_2^2}{2 V^3} a^2 k_0^2 n_2 A_m (A_{m-1} - A_{m+1})  D_m D_{m+1} d \nonumber \\
        &- \frac{n_1^2 + n_2^2}{V} a^2 k_0^2 n_2  A_m (A_{m-1} - A_{m+1}) (D_m D_{m-1}  + D_m F_{m+1}  + D_{m+1} F_m ) d \nonumber \\
       &- \frac{n_1^2 + n_2^2}{2 V} \left[ A_m (B_{m-1} - B_{m+1})  + B_m ( A_{m-1} - A_{m+1} ) \right] D_m D_{m+1} d \nonumber
\end{align}

The whole determinant is thus 
\begin{align}
    \frac{2 (a q)^{2m + 2}}{a^2 k_0^2} & \det M 
  \approx  \label{suppl_eq:wholedet} \\
       &      n_2^2 A_m^2 D_{m+1} (D_{m-1} + F_{m+1})  
         -   4 m^2 n_2^2 A_m^2 D_m^2 \frac{1}{V^2}   \nonumber \\
         -   & 2 \frac{m^2}{(a k_0)^2} A_m^2 D_m^2
         - 4 m^2 n_2^2 A_m^2 D_m F_m   \nonumber \\
       - &\frac{n_1^2 + n_2^2}{2 V} A_m (A_{m-1} - A_{m+1})  D_m D_{m+1}   \nonumber \\
       + &~   \nonumber \\
       & \left\{ a^2 k_0^2 n_2^3 A_m^2 \left[2 D_{m+1} F_{m-1} + 2 D_{m+1} G_{m+1} + (D_{m-1} + F_{m+1})^2 \right]  \right.   \nonumber \\
       -&   4 a^2 k_0^2 m^2 n_2^3 A_m^2 \left[ 3 \frac{D_m^2}{V^4}   +   F_m^2 + 2 D_m G_m \right]     \nonumber \\ 
       -&    8 m^2 n_2 A_m^2 D_m \left[ \frac{D_m }{V^2}  +  2 \frac{ a^2 k_0^2 n_2^2 F_m }{V^2} + F_m \right] \nonumber   \\
         +&   2 n_2^2 A_m B_m D_{m+1} (D_{m-1} + F_{m+1})  
         -   8 m^2 n_2^2 A_m B_m D_m^2 \frac{1}{V^2}   \nonumber \\
         -&    4 \frac{m^2}{(a k_0)^2} A_m B_m D_m^2  
         - 8 m^2 n_2^2 A_m B_m D_m F_m    \nonumber \\
        +& \frac{(a k_0)^2 n_1^2 n_2}{V^2} (  A_{m-1}^2 - 2 A_{m-1} A_{m+1} + A_{m+1}^2  ) D_m^2    \nonumber \\ 
       -& \frac{n_1^2 + n_2^2}{2 V^3} a^2 k_0^2 n_2 A_m (A_{m-1} - A_{m+1})  D_m D_{m+1}     \nonumber \\ 
        -& \frac{n_1^2 + n_2^2}{V} a^2 k_0^2 n_2  A_m (A_{m-1} - A_{m+1}) (D_m D_{m-1}  + D_m F_{m+1}  + D_{m+1} F_m )    \nonumber \\ 
       -& \left. \frac{n_1^2 + n_2^2}{2 V} \left[ A_m (B_{m-1} - B_{m+1})  + B_m ( A_{m-1} - A_{m+1} ) \right] D_m D_{m+1} \right\} d   \nonumber \\    
       \stackrel{\text{def}}{=} &a_{M,m} + b_{M,m} d \nonumber
\end{align}
The RHS of the equation is of the form $a_{M,m} + b_{M,m} d$, where $a_{M,m}$ and $b_{M,m}$ are constants that depend on the wavelength, azimuthal order $m$, and the fiber parameters $a$, $n_1$, and $n_2$. $a_M$ is given by the sum of the terms above the plus sign in its own line, and $b_M$ is all the $d$-dependent terms given by the bracketed expression below the plus sign. We can also see that the factor in front of the determinant is zero if and only if $q = 0$, and $q = 0$ only at the cutoff wavelengths. Outside of the cutoff, the determinant is thus zero if and only if the right-hand side of the equation equals zero, and $d$ can now be solved from this condition. For $m \geq 3$ we have
\begin{equation}
    d = - \frac{a_{M,m}}{b_{M,m}} ,
\end{equation}
but for $m \leq 2$ the coefficients may involve logarithms of $d$, and the equation is not as easy to solve, and these cases would have to be treated separately.

\subsection{The Numerator $a_{M,m}$}

The expression for the term $a_{M,m}$ for $m \geq 1$ is
\begin{align}
  a_{M,m} =  
       &n_2^2 A_m^2 D_{m+1} (D_{m-1} + F_{m+1})  
         -   4 m^2 n_2^2 A_m^2 D_m^2 \frac{1}{V^2}   \nonumber \\
         -   & 2 \frac{m^2}{(a k_0)^2} A_m^2 D_m^2
         - 4 m^2 n_2^2 A_m^2 D_m F_m  
       - \frac{n_1^2 + n_2^2}{2 V} A_m (A_{m-1} - A_{m+1})  D_m D_{m+1}  .
\end{align}
We can first use the identities  $D_{m + 1} = 2 m D_m$, $2 m J_m(V) = V[ J_{m-1}(V) + J_{m+1}(V)]$, and $a/(a k_0)^2 = (n_1^2 - n_2^2)/V^2$ to get
\begin{align}
  a_{M,m} =
       &2 m n_2^2 A_m^2 D_m ( D_{m-1} + F_{m+1} )  
         -   4 m^2 n_2^2 A_m^2 D_m^2 \frac{1}{V^2}   \nonumber \\
         -   & 2 \frac{m^2 (n_1^2 - n_2^2) }{V^2} A_m^2 D_m^2
         - 4 m^2 n_2^2 A_m^2 D_m F_m  
       - 2 m \frac{n_1^2 + n_2^2}{V^2} A_m (V A_{m-1} - m A_m)  D_m^2   ,
\end{align}
which simplifies to
\begin{equation}
  a_{M,m} =     2 m A_m D_m 
  \left[  n_2^2 A_m ( D_{m-1} + F_{m+1} ) - 2 m n_2^2 A_m F_m  - \frac{n_1^2 + n_2^2}{V} A_{m-1} D_m \right] .
\end{equation}
For $m = 1$ we have
\begin{align}
  a_{M,1} =    & 2 J_1 D_1 
  \left[  n_2^2 J_1 ( D_0 + F_2 ) - 2 n_2^2 A_1 F_1  - \frac{n_1^2 + n_2^2}{V} J_0 D_1 \right]  \nonumber \\
          =    & 2 J_1 
  \left\{  n_2^2 J_1 \left[ \ln 2 - \gamma - \ln (aq)  - \frac{1}{2} \right] - n_2^2 J_1 \left[ \gamma - \frac{1}{2} - \ln 2 + \ln (a q)  \right]  - \frac{n_1^2 + n_2^2}{V} J_0 \right\}  \nonumber \\
    =    & 2 J_1 
  \left\{  2 n_2^2 J_1 \left[ \ln 2 - \gamma - \ln (aq)   \right] - \frac{n_1^2 + n_2^2}{V} J_0 \right\}  .
\end{align}
Near cutoff the logarithm becomes very large, which shows that $a_{M,1}$ can only be zero if $J_1(V) = 0$. Thus, this is the only cutoff condition for modes of azimuthal order $m = 1$, and HE- and EH-modes have the same cutoff wavelengths. Given that near cutoff the logarithm becomes dominant over the constant terms, we have
\begin{equation}
    a_{M,1} \approx - 4 n_2^2 J_1^2(V)  \ln (aq) .
\end{equation}
For $m\geq 2$ the expression for $a_{M,m}$ does not contain logarithms, and it simplifies to 
\begin{equation}
  a_{M,m} =     2^{2m - 1} m [(m - 1)!]^2 J_m(V) 
  \left[ \frac{n_2^2}{m - 1} J_m(V)  - \frac{n_1^2 + n_2^2}{V} J_{m-1}(V)   \right] . 
\end{equation}
We can readily see that now $a_{M,m}$ is zero if and only if either $J_m(V) = 0$ or $V n_2^2 J_m(V)  - (m - 1) (n_1^2 + n_2^2) J_{m-1}(V) = 0$. These two conditions correspond to the cutoff conditions for EH- and HE-modes, respectively, and it is evident that the cutoff wavelengths for HE- and EH-modes are now different, as both conditions cannot be fulfilled simultaneously (due to the property of Bessel functions that $J_m(V) = 0$ directly implies $J_{m-1}(V) \neq 0$).

\subsection{The Denominator $b_{M,m}$ and the Approximate Effective Index Near Cutoff}

The general, full expression for $b_{M,m}$ is
\begin{align}
b_{M,m} = 
     &a^2 k_0^2 n_2^3 A_m^2 \left[2 D_{m+1} F_{m-1} + 2 D_{m+1} G_{m+1} + (D_{m-1} + F_{m+1})^2 \right]     \nonumber \\
       -&   4 a^2 k_0^2 m^2 n_2^3 A_m^2 \left[ 3 \frac{D_m^2}{V^4}   +   F_m^2 + 2 D_m G_m \right]     \nonumber \\ 
       -&    8 m^2 n_2 A_m^2 D_m \left[ \frac{D_m }{V^2}  +  2 \frac{ a^2 k_0^2 n_2^2 F_m }{V^2} + F_m \right]   \label{suppl_eq:wholedetb} \\
         +&   2 n_2^2 A_m B_m D_{m+1} (D_{m-1} + F_{m+1})  
         -   8 m^2 n_2^2 A_m B_m D_m^2 \frac{1}{V^2}   \nonumber \\
         -&    4 \frac{m^2}{(a k_0)^2} A_m B_m D_m^2  
         - 8 m^2 n_2^2 A_m B_m D_m F_m    \nonumber \\
        +& \frac{(a k_0)^2 n_1^2 n_2}{V^2} (  A_{m-1}^2 - 2 A_{m-1} A_{m+1} + A_{m+1}^2  ) D_m^2    \nonumber \\ 
       -& \frac{n_1^2 + n_2^2}{2 V^3} a^2 k_0^2 n_2 A_m (A_{m-1} - A_{m+1})  D_m D_{m+1}     \nonumber \\ 
        -& \frac{n_1^2 + n_2^2}{V} a^2 k_0^2 n_2  A_m (A_{m-1} - A_{m+1}) (D_m D_{m-1}  + D_m F_{m+1}  + D_{m+1} F_m )    \nonumber \\ 
       -& \frac{n_1^2 + n_2^2}{2 V} \left[ A_m (B_{m-1} - B_{m+1})  + B_m ( A_{m-1} - A_{m+1} ) \right] D_m D_{m+1} .
\end{align}
For $m = 1$ the expression for $b_{M,1}$ involves the logarithm $\ln(aq)$ just like $a_{M,1}$ does. these are the dominant terms. There are also terms proportional to $\ln(aq)^2$ from terms such as $F_1^2$, but these cancel each other out. For small $d$, the expression for $b_{M,1}$ becomes
\begin{equation}
    b_{M,1} \approx  - 4 n_2 J_1(V) \left\{ 2 \frac{a^2 k_0^2 n_2^2 }{V^2}  J_1(V) + J_1(V) 
    - \frac{a^2 k_0^2 n_2^2}{V} [J_0(V) - J_2(V)]   \right\} \ln(a q) .
\end{equation}
The approximate effective index thus becomes
\begin{align}
    \neff &\approx n_2 - \frac{a_{M,1}}{b_{M,1}} \\
          &\approx n_2 - \frac{n_2 V^2 J_1(V) }{ 2 a^2 k_0^2 n_2^2 J_1(V) + V^2 J_1(V) - a^2 k_0^2 n_2^2 V [J_0(V) - J_2(V)]   } .
\end{align}
We note that the approximation above relies on the logarithm terms becoming dominant, which requires the effective index to be very close to the cladding index. Better approximations could be obtained by keeping all the terms in $a_{M,m}$, $b_{M,m}$, and solving the equation $a_{M,m} + b_{M,m} d = 0$, the solutions of which can be expressed in closed form using the Lambert $W$-function.

The expression for $b_{M,2}$ also involves terms proportional to $\ln(aq)$, which are the ones that become dominant close to the cutoff wavelength. However, $A_{M,2}$ contains no logarithms. Only keeping these dominant terms yields
\begin{equation}
    b_{M,2} \approx 16 J_2(V)^2 a^2 k_0^2 n_2^3 \ln(a q) 
           \approx 8 J_2(V)^2 a^2 k_0^2 n_2^3 \ln d .
\end{equation}
The determinant equation thus becomes 
\begin{align}
    a_{M,2} + b_{M,2} d &= 0 \nonumber \\
    16 J_2(V) \left[ n_2^2 J_2(V)  - \frac{n_1^2 + n_2^2}{V} J_1(V)   \right]
    + 8 J_2(V)^2 a^2 k_0^2 n_2^3 d \ln d &= 0 \nonumber \\
    2 \left[ n_2^2 J_2(V)  - \frac{n_1^2 + n_2^2}{V} J_1(V)   \right]
    + J_2(V) a^2 k_0^2 n_2^3 d \ln d &= 0 ,
\end{align}
which, again, can be solved in terms of the Lambert $W$-function but hence will not yield an evident computational advantage. Furthermore, this approximation only works for HE-type modes, as $J_m(V) = 0$ for EH-type modes, which causes the solutions to the equation above to diverge. This is a manifestation of terms already neglected along the way actually becoming important again. Similar expressions for the TE- and TM-modes can also be derived, but they involve even more logarithms due to the associated lower orders of the Bessel functions and will not be considered here, though the same princples can be applied to these modes as well.

The most general case of azimuthal order $m \geq 3$ is luckily the easiest to deal with, as it lacks the logarithm terms present in the approximations for smaller azimuthal mode orders. For $m \geq 3$ all the factors in the terms for $a_{M,m}$ and $b_{M,m}$ are either constants or Bessel functions, all of which can be written in terms of $J_m(V)$ and $J_{m+1}(V)$. Starting with the general expression for $b_{M,m}$, we can first write $D_{m-1}, D_{m+1}, F_{m-1}$ etc. in terms of $D_m, F_m,$ and $G_m$. All the remaining Bessel functions can be written in terms of $J_m(V)$ and $J_{m+1}(V)$, and the expression for $b_{M,b}$ simplifies to
\begin{align}
b_{M,m} = 
     &    \frac{2^{2m - 1} [(m-1)!]^2 }{V^4 (m - 1)} n_2 [   S_1 J_{m+1}^2 +  S_2 V J_{m+1} J_m  +  S_3 J_{m}^2 ] ,
\end{align}
where
\begin{align}
    S_1 &= (m + 2)(m-1)  V^4  + 2 f (m^2 - 1) V^2 \\
    S_2 &= - (m - 1) V^4 - 2 (2 m^3 - m^2 - m - f) V^2 - 8 (m - 1) f m^2 \\
    S_3 &=   \frac{m^3 - 2 m^2 - f m + f}{m - 2} V^4 + 4 m^2 (m - 1)^2 V^2 + 8 f m^2 (m - 1)^2 ,
\end{align}
and $f = a^2 k_0^2 n_2^2$.
The effective index for $m \geq 3$ is thus
\begin{align}
    \neff \approx \neff^\text{(appr.)} = n_2 - \frac{m J_m}{n_2} \frac{  V n_2^2 J_m  - (m - 1) (n_1^2 + n_2^2) J_{m-1}}
    { S_1 J_{m+1}^2 +  S_2 V J_{m+1} J_m +  S_3 J_{m}^2 } V^3 . \label{suppl_eq:neffappr}
\end{align}
As a reminder, the argument for $J_m$ and $J_{m+1}$ that has been left out is the normalized frequency $V = a k_0 \sqrt{n_1^2 - n_2^2}$.

\section{Linear Approximation to the Effective Index Near Cutoff with Respect to Wavelength}

The RHS of Eq.~(\ref{suppl_eq:neffappr}) is clearly nonlinear in the vacuum wavelength $\lambda_0$, as the wavelength appears inside $V$ and inside Bessel functions. This nonlinear behavior generally causes the approximation given by Eq.~(\ref{suppl_eq:neffappr}) to become poor quite rapidly when the wavelength is not close to the cutoff wavelength. However, the approximation was done to first order in $d$ and hence the actual effective index and its approximation are locally co-linear at the cutoff wavelength as a function of wavelength. The approximation of Eq.~(\ref{suppl_eq:neffappr}) can therefore be utilized to derive linear approximations of the form $\neff \approx n_2 + \kappa (\lambda_0 - \lambda_c)$, where $\kappa$ is a constant and $\lambda_c$ is a cutoff wavelength of the mode of interest. At the cutoff wavelength, the mode effective index becomes equal to the cladding index. Equation~(\ref{suppl_eq:neffappr}) then gives this cutoff condition as 
\begin{align}
     J_m   \left[  V n_2^2 J_m  - (m - 1) (n_1^2 + n_2^2) J_{m-1}  \right]  = 0. \label{suppl_eq:cutoffcond}
\end{align}
Indeed, this equation reflects the well-known cutoff conditions for modes:
\begin{equation}
    J_m(V_c) = 0 \label{suppl_eq:EHcutoffcond}
\end{equation}
for EH-modes and
\begin{equation}
    V_c n_2^2 J_m(V_c)  - (m - 1) (n_1^2 + n_2^2) J_{m-1}(V_c) = 0 \label{suppl_eq:HEcutoffcond}
\end{equation}
for HE modes, where
\begin{equation}
    V_c = \frac{2 \pi a }{\lambda_c} \sqrt{n_1^2 - n_2^2} .
\end{equation}
The cutoff wavelengths $\lambda_c$ can be solved numerically from these conditions. Let us denote the solutions of Eq.~(\ref{suppl_eq:EHcutoffcond}) in ascending order by $j_{mn}$, as is conventional. These are just the zeros of the Bessel function $J_m$. Let $s_{mn}$ denote the solutions of Eq.~(\ref{suppl_eq:HEcutoffcond}) in ascending order ($s_{mn} < s_{m(n+1)}$) .

The get an approximation of the form $\neff \approx n_2 + \kappa (\lambda_0 - \lambda_c)$ we simply differentiate $\nappr$ in Eq.~(\ref{suppl_eq:neffappr}) with respect to the wavelength and evaluate the derivative at $\lambda_c$ to get $\kappa$. It is beneficial to use the chain rule as
\begin{equation}
    \frac{d \neff^{(\text{appr.})}}{d \lambda_0} 
    =
    \frac{d V}{d \lambda_0}  \frac{d \neff^{(\text{appr.})}}{d V} 
    =
    - \frac{V}{\lambda_0}  \frac{d \neff^{(\text{appr.})}}{d V}    . \label{suppl_eq:chain}
\end{equation}
The product rule also comes in handy when used in the following manner:
\begin{align}
    \frac{d \neff^{(\text{appr.})}}{d V} 
    &= 
    \frac{m}{n_2}  
\left[ \frac{  V n_2^2 J_m  - (m - 1) (n_1^2 + n_2^2) J_{m-1}}
    { S_1 J_{m+1}^2 +  S_2 V J_{m+1} J_m +  S_3 J_{m}^2 } V^3 \right] \frac{d J_m}{d V} \nonumber \\
&+ \frac{m}{n_2} J_m \frac{d}{d V}\left[ \frac{  V n_2^2 J_m  - (m - 1) (n_1^2 + n_2^2) J_{m-1}}
    { S_1 J_{m+1}^2 +  S_2 V J_{m+1} J_m +  S_3 J_{m}^2 } V^3 \right] \label{suppl_eq:neffdiff} .
\end{align}
The reason for dividing the terms like this is that the top term is zero (at $V_c$) for HE modes and the bottom term is zero for EH modes due to their cutoff conditions. Also note that both cutoff conditions together with the Bessel function recursion relations allow for $J_m(V_c)$ and $J_{m-1}(V_c)$ to be written solely as $J_{m+1}(V_c)$ times a constant. This means that both the numerator and the denominator contain factors of $J_{m+1}^2$, and they cancel out. Calculating the derivatives yields
\begin{align}
    \neff(\text{HE}_{mn}) &\approx 
    n_2 - \frac{\lambda_0 - \lambda_c}{\lambda_c} \frac{m (n_1^4 - n_2^4)}{n_2}  \frac{(m-1)^2 (n_1^2 + n_2^2) + n_2^2 g }{P_m + Q_m g}  \label{suppl_eq:HEneffc} \\
    &=
    n_2 - \left( \frac{s_{mn}}{2 \pi a \sqrt{n_1^2 - n_2^2}} \lambda_0 - 1 \right) \frac{m (n_1^4 - n_2^4)}{n_2}
    \frac{(m-1)^2 (n_1^4 - n_2^4) + n_2^4 s_{mn}^2}{ (n_1^2 - n_2^2) P_m + n_2^2 Q_m s_{mn}^2 }  \label{suppl_eq:HEneff} .
\end{align}
where $g = (2 \pi a n_2 / \lambda_c)^2$ and
\begin{equation}
    P_m = m (m-1) (m-2) n_1^2 + m^2 (m-1) n_2^2
\end{equation}
and
\begin{equation}
    Q_m = \frac{m-1}{m-2} (n_1^2 + n_2^2)^2  + (m-2) n_2^2 (n_1^2 + n_2^2) + 2 n_2^4 
\end{equation}
for HE modes. For EH modes we get
\begin{align}
    \neff(\text{EH}_{mn}) 
    &\approx n_2 - \frac{\lambda_0 - \lambda_c}{\lambda_c}
              \frac{m (n_1^4 - n_2^4)}{n_2 [  (m + 2) n_1^2 + m n_2^2 ]} \label{suppl_eq:EHneffc} \\
    &=  n_2 - \left( \frac{j_{mn}}{2 \pi a \sqrt{n_1^2 - n_2^2}} \lambda_0 - 1 \right)
              \frac{m (n_1^4 - n_2^4) }{n_2 [  (m + 2) n_1^2 + m n_2^2 ]} \label{suppl_eq:EHneff} .
\end{align}

\end{document}